\documentclass[12pt]{article}

\usepackage{graphicx}
\usepackage{setspace}
\usepackage{fancybox}
\usepackage{pseudocode}
\usepackage[margin=1.2in]{geometry}

\begin{document}

\title{Spatio-Temporal Disease Surveillance: Forward Selection Scan Statistic}
\author{Ross Sparks and Adrien Ickowicz\\
CSIRO Mathematics, Informatics and Statistics\\ Private Bag 17\\ North Ryde\\ NSW 1670 \\ Australia\\
\emph{e-mail}:Ross.Sparks@csiro.au\\}

\maketitle

\begin{abstract}
 The scan statistic sets the benchmark for spatio-temporal surveillance methods with its popularity.  In its simplest form it scans the target area and time to find regions with disease count higher than expected.  If the shape and size of the disease outbreaks are known, then to detect it sufficiently early the scan statistic can design its search area to be efficient for this shape and size. A plan that is efficient at detecting a range of disease outbreak shapes and sizes is important because these vary from one outbreak to the next and are generally never known in advance.  This paper offers a forward selection scan statistic that reduces the computational effort on the usual single window scan plan, while still offering greater flexibility in signalling outbreaks of varying shapes.  The approach starts by dividing the target geographical regions into a lattice. Secondly it smooths the time series of lattice cell counts using multivariate exponential weighted moving averages.  Thirdly, these EWMA cell counts are spatially smoothed to reduce spatial noise and leave the spatial signal. The fourth step uses forward selection approach to scanning mutually exclusive and exhaustive rectangular regions of dynamic  dimensions. In the fifth step, it prunes away all insignificant scanned regions where counts are  not significantly higher than expected. An outbreak is signalled if at least one region remains after pruning. If all regions are pruned away - including the scan of the target region, then no outbreak is signalled.

\end{abstract}
	

\section{Introduction}

The Scan statistic (Kulldorff, 2001) has been proposed for detecting space-time disease clusters, but it is computational intensive and therefore difficult to scale up to higher dimensions. For example, the dimensions may include residential and work address, severity, age groups, symptoms etc.  Kulldorff (1995, 1997, and 2005) developed scan plans and implemented them in the SATSCAN software package for a variety of problems including Bernoulli data, Poisson counts and a space-time permutation model using only case data, amongst others.  The important limitations to scan statistic are:
\begin{enumerate}
  \item The space-time permutation model assumes that all cases were independent of each other.  However in practice common cause variation generally involves significant space-time interactions, e.g., seasonal influences may vary across the target geographical region.
  \item The spatio-temporal scan statistic has been criticised by Woodall et al. (2008) and Han et al. (2008) for not being as efficient as the CUSUM (see Raubertas, 1989, and Rogerson and Yamada, 2004) for outbreak detection.
  \item The ability to detect outbreaks most effectively is dependent on the choice of shape and size of the scanning window.
  \item The approach is not easy to extend to higher dimensions.
\end{enumerate}
\noindent The attractiveness of the scan statistic is in its conceptional simplicity.  There is a dynamic size window scan plan in the literature (Takahashi {\it et al.}, 2008) but this is only applicable if the population sizes within each geographical cell is known.

The two dimensional scan statistic used for detecting spatial clusters (Glaz {\it et al.},2001) is the approach that will be compared with the FSS plan in this paper.  The advantage of this approach is the ease in applying the multivariate EWMA to accumulate memory of past cell counts like Grigg and Spiegelhalter (2007) did in their paper.

The forward selection scan (FSS) plan proposed addresses all of the concerns raised above, by first applying a spatio-temporal EWMA smoothing for the cell counts before using a flexible forward selection approach to choosing the scanned region.  Outbreaks are signalled in this paper when the counts are significantly higher than expected in a rectangular region of the target geography.  Section 2 of the paper introduces the single size window scan statistic. Section 3 describes the FSS plan.  Section 4 provides a simulation study for comparing plans under a number of different outbreak scenarios where Average Run Length (ARL) is used to assess and compare the detection properties of the plans. Section 5 briefly covers an example of application
relating to gastrointestinal disease in the county of Hampshire, UK (see Diggle, {\it et al.}, 2003).

\section{The Scan Statistic}

The scan statistic is a spatio-temporal plan that counts the number of observations in a regular prism of spatio-temporal space. The height of the prism is time (taken as days in this paper) which represents temporal memory in the plan. This paper scans all possible prisms of the same dimensions in the search for outbreaks. Prism counts are compared to their respective expected counts to gauge their unusualness.  Counts much higher than expected signal outbreaks.  The ideal design is to have all scanned blocks with roughly the same expected values; however this is difficult in practice.  Blocks in this paper are therefore constructed so that the marginal row/column totals have the same expected values. This is not necessary but does give each column and row the same power of detection for an outbreak of the same size.

For the target region involving a spatial lattice with $A$ rows and $B$ columns, let the daily disease cases for the cell in the  $i$th row and $j$th column in the lattice in day $t$  be recorded as the random variable $Y_{ijt}$. Let  ${\rm E}(Y_{ijt})=\mu_{ijt}$.   The total cases in a scanned spatio-temporal window of size  $m_1 \times m_2 \times T$ for starting cell with row $i$ and column $j$  is
	 	
\begin{equation}\label{(1)}
  \underline{Y}_{ijt}=\sum_{\tau=t-T+1}^{t}\sum_{\ell=i}^{i+m_1-1}\sum_{k=j}^{i+m_2-1}Y_{ \ell k \tau}
\end{equation}

with expected values
\begin{equation}\label{(1)}
  M_{ijt}=\sum_{\tau=t-T+1}^{t}\sum_{\ell=i}^{i+m_1-1}\sum_{k=j}^{i+m_2-1}\mu_{\ell k \tau}
\end{equation}

\noindent where $i=1,2,\dots,A-m_1+1$, $j=1,2,\dots,B-m_2+1$.

The counts $\underline{Y}_{ijt}$ are assumed to be Poisson distributed with mean $M_{ijt}$.  Thus Poisson tables are used in testing whether  $\underline{Y}_{ijt}$ values are significantly higher than $M_{ijt}$.  Since we are testing whether $\underline{Y}_{ijt}$ values are significantly larger than expected for $i=1,2,…,A-m+1$, $j=1,2,…,B-m+1$ simultaneously, we need to adjust the level of significance for this multiple testing.

For observed count $\underline{y}_{ijt}$, a signal is given whenever
$$p_{ijt}=P(\underline{Y}_{ijt}< \underline{y}_{ijt} \vert M_{ijt})>h_{scan}$$
for at least one $i=1,2,\dots,A-m_1+1$, $j=1,2,\dots,B-m_2+1$. The value of  $h_{scan}$ is a threshold designed to give a fixed in-control ARL. Starting values for $h_{scan}$ can be found by first aggregating over time and then use Chen and Glaz (1996).  Simulation is then used to refine this threshold to deliver the appropriate in-control ARL.

This Scan plan is applied in this paper with the following parameters:
\begin{itemize}
  \item A target block with $A=40$ and $B=40$
  \item The depth of window (time) is arbitrarily taken as $T=10$.
  \item The spatial scan has $m_1=m_2=10$, the width and length of the window are equal to 10 cells.
\end{itemize}

\section{Forward Selection Scan (FSS) Plan}

\subsection{General overview}

The FSS plan proposed here will be compared to the single size window scan plan discussed in Section 2.  FSS plan consists of four major steps:\begin{enumerate}
 \item {\bf Multivariate EWMA step:} There are two ways of viewing this step. Firstly, the cell counts are temporally smoothed as follows $\bar{y}_{ijt}=\alpha y_{ijt}+(1-\alpha) \bar{y}_{ijt-1}$ for all $i,j$ and $t$($>0$) with $\bar{y}_{ij0}=\mu_{ij1}$ and $0<\alpha<1$ (see Lowry et al. 1992, Sparks, 1992 for multivariate EWMA of continuous data, Sparks et al., 2010 for non-homogeneous Poisson counts) leaving mostly cell trends. Similarly the FSS plan smooths the expected counts $\bar{\mu}_{ijt}=\alpha \mu_{ijt}+(1-\alpha) \bar{\mu}_{ijt-1}$ in the same way.  The second interpretation of this step is that the EWMA accumulates memory of past counts with an exponential decaying memory.
 \item {\bf Spatial Smoothing step:}  Here we define matrix $\bar{Y}_t=\{\bar{y}_{ijt}\}$, and $\bar{M}_t=\{\bar{\mu}_{ijt}\}$. Let $S_A$ and $S_B$ be $A \times A$ and $B \times B$ symmetric smoothing matrices such that $S_X=\{s_{ij}=\lambda(1-\lambda)^{|i-j|}\}$ where $X=A,B$, and $ 0 < \lambda < 1$. Let $1_A$ be a vector of $A$ ones,  $D_A$ is a diagonal matrix with elements equal to $S_A \, 1_A$ and $D_B$ is a diagonal matrix with elements equal to $S_B\, 1_B$.  Then the spatial smoothed of the MEWMA counts is given by $\tilde{Y}_t=D_A^{-1}S_A\bar{Y}_t S_B D_B^{-1}$. Similarly the smoothed means are $\tilde{M}_t=D_A^{-1}S_A \bar{M}_t S_B D_B^{-1}$.  This is similar to using a kernel smoother involving independent double exponential distributions for rows and columns.
 \item {\bf Forward selection of partitions step:} Find longitude partitions or latitude partitions that divide the parent region into two offspring (mutually exclusive and exhaustive sub-spaces) such that smoothed counts for one offspring region has the most unusually higher than expected smoothed count.  This binary recursive partitioning approach will be outlined in more detail in the next subsection.
 \item {\bf Prune away insignificant offspring step:} This step is designed to prune away all regions generated in the previous step that do not depart significantly from their expected values.  Each generation of offspring are pruned away recursively if they fail to exceed a significance threshold.  If all generations are pruned then no outbreak is signalled, otherwise there is a signal.
 \end{enumerate}

The level of significance of outbreaks ($p-$values) cannot be used to determine the best partition because it often hits the boundary of zero.  Therefore some signal-to-noise ratio is required to replace it.  The usual signal-to-noise ratio of counts minus mean, divided by standard deviation can suffer instabilities when the mean gets very small.  An alternative \lq signal-to-noise\rq, \ after counts have been spatially and temporally smoothed, is
$$\sqrt{\tilde{Y}_{ijt}}-\sqrt{\tilde{M}_{ijt}}$$
which is approximately normally distributed with mean zero and roughly constant variance (for all $\tilde{M}_{ijt}$).\\

\subsection{Partitioning Step}

FSS plan generates offspring by recursively partitioning either longitudinal cells or latitudinal cells into rectangular regions.   The process begins with the whole target area. Each partition divides the parent space into two mutually exclusive and exhaustive sub-spaces(called offspring), so that one of the offspring has the most unusually high smoothed disease counts.  The term unusually high can be interpreted as the smoothed count with the lowest p-value given its expected distribution.  However this approach would not adjust for the degrees of freedom in the partition process such as the variable amount of longitude and latitude searching.  This issue will be discussed later.

Each partition of a parent region results in two offspring, one region with the most unusually high smoothed count and the other is the remainder of the parent region.  The process keeps growing each new generation until a stopping rule is reached.  The stopping rule (discussed later) terminates generating further offspring for that parent. Once partitioning has stopped for all generations, then pruning of the offspring regions commences.  This pruning process is outlined in the next section.

A measure of how far the smoothed count departs from the expected is needed.  We use the robust measure of the square root of the regional sum of smoothed counts minus the square root of the regional sum of smoothed expected counts. This is very close to normally distributed, and when in-control it has an expected value of zero (and approximately constant variance).

Let the parent space involve rows $i, i+1, \dots,i+m_1-1$ and columns $j,j+1,\dots,j+m_2-1$. The recursive partitioning process for finding outbreaks for this parent space involves the following steps:
\begin{enumerate}
  \item {\bf Find the best row partitions:} Find the $k$ which maximises either $$R_{1k}=\sqrt{\sum_{n=i}^{i+k} ~ \sum_{\ell=j}^{j+m_2-1} \tilde{Y}_{n \ell t}} \ - \ \sqrt{\sum_{n=i}^{i+k} ~ \sum_{\ell=j}^{j+m_2-1} \tilde{M}_{n \ell t}}$$ or
      $$R_{2k}=\sqrt{\sum_{n=i+k+1}^{i+m_1-1} ~ \sum_{\ell=j}^{j+m_2-1} \tilde{Y}_{n \ell t}}\ - \ \sqrt{\sum_{n=i+k+1}^{i+m_1-1} ~ \sum_{\ell=j}^{j+m_2-1} \tilde{M}_{n  \ell t}}$$ for $k=0,1,\dots,m_1-1$.  
      The most unusual row partition corresponds to $R_k^{\max}=\max(R_{1k},R_{2k})$.
  \item {\bf Find the best column partitions:} Find the $k$ which maximises either $$C_{1k}=\sqrt{\sum_{n=i}^{i+m_1-1} ~ \sum_{\ell=j}^{j+k} \tilde{Y}_{n \ell t}}\ - \ \sqrt{\sum_{n=i}^{i+m_1-1} ~ \sum_{\ell=j}^{j+k} \tilde{M}_{n \ell t}}$$ or
      $$C_{2k}=\sqrt{\sum_{n=i}^{i+m_1-1} ~ \sum_{\ell=j+k+1}^{j+m_2-1} \tilde{Y}_{n \ell t}}\ - \ \sqrt{\sum_{n=i}^{i+m_1-1} ~ \sum_{\ell=j+k+1}^{j+m_2-1} \tilde{M}_{n \ell t}}$$ for $k=0,1,\dots,m_2-1$.The most unusual column partition corresponds to $C_k^{\max}=\max(C_{1k},C_{2k})$.
  \item {\bf Decide whether to partition on rows or columns:} We could partition on the rows if $R_k^{\max}>C_k^{\max}$ and on column otherwise, and let $P_k=\max(R_k^{\max},C_k^{\max})$ as discussed later. However, this does not adjust for the degrees of freedom in the amount of variable searching along the x- and y-coordinates of the parent space (columns and rows).  For example, if the parent space spanned 40 by 2 cells in the lattice, then, when in control, the partition using ${\rm max}(R_k^{\max},C_k^{\max})$ is more likely along the 40 rows than the two columns.  Not conditioning on the offspring degree of freedom for the counts fails to find the most unusual shift in counts from expected.  For example, if the child expected count is very close to the parent expected count then there is little degree of freedom for the offspring count to improve on the signal-to-noise ratio of the parent.  Therefore parent and offspring expected counts both influences the degree of freedom of movement in the offspring signal-to-noise ratio.  The variation in parent counts also influences the degree of freedom of movement in the child counts because the child counts is less than or equal to the parent count. In this paper we want to partition along the dimension that is most unusual after conditioning on the aspects that relate to the offspring degrees of freedom to move from expected.  These aspects are:
      \begin{enumerate}
       \item The amount of searching - the number of potential partitions, e.g., in the 40 by 2 parent space discussed above it is 39 row partitions and one column partition.
       \item The parent counts and parent mean counts - the more the parent count departs from its expected the easier it is to generate a signal-to-noise ratio that is high for an offspring.
       \item The offspring mean count - the larger this offspring mean is relative to the parent mean counts, the less we expect in-control offspring counts to vary from its expected value.
             \end{enumerate}
       Therefore we wish to find the expected value and variance of $P_k$ conditional on the degrees of freedom aspects mention in (a) to (c) above.  The next step is to estimate this conditional expectation. A parametric bootstrap approach can be used.
Parametric bootstrap samples of in-control timely counts are generated.  The population mean values are known for each cell.  We smooth these counts spatially and temporal as described in the beginning of section 3.  These smoothed counts are used to find partitions by splitting on the rows if $R_k^{\max}>C_k^{\max}$ and on column otherwise.  We repeat this for all of in-count generated counts giving an upper bound of partitions with $R_k^{\max}$ and $C_k^{\max}$ values.  For each partition we record both $R_k^{\max}$ and $C_k^{\max}$ values and the following corresponding values
       \begin{itemize}
      \item $n_s$ the number of possible partitions there where in the parent space for each $R_k^{\max}$ and $C_k^{\max}$, respectively.
          \item $\mu={\rm E}(R_{jk})$ corresponding to $R_k^{\max}$ and $\mu={\rm E}(C_{jk})$ corresponding to $C_k^{\max}$.
              \item $\mu_p$ is the parent space expected counts and $c_p$ the count for the parent region.
              \item $z_p=2(\sqrt{c_p}-\sqrt{\mu_p})$.
              \end{itemize}
          Although, the values of $\mu_p$ and $z_p$ are common for both row and column partitions for each offspring from the same parent, and therefore their values do not influence the selection; it is helpful to condition on these to build a good model because they do related to offspring degree of freedom.  These recorded values are used then to fit a model to predict ${\rm E}(X\vert n_s,\mu,\mu_p,c_p)$ (denoted ${\rm E}(X)$ from now on) where $X=R_k^{\max}$ or $C_k^{\max}$.  Similarly we find the condition variance and denote this ${\rm Var}(X)$.

  \item {\bf Repeat recursively steps 1, 2 and 3 above until a stopping rule is applied:} For each new generation repeat the process for the two offspring generated using the steps above.  If we partitioned on the rows then the next generation parent spaces are defined by
      \begin{itemize}
      \item rows $i,i+1,\dots,k$ and columns $j,j+1,\dots,j+m_2-1$
      \item rows $i+k+1,i+k+2,\dots,i+m_1-1$ and columns $j,j+1,\dots,j+m_2-1$
      \end{itemize}
      If we partitioned on the columns then the next generation parent spaces are defined by
      \begin{itemize}
      \item rows $i,i+1,\dots,i+m_1-1$ and columns $j,j+1,\dots,j+k$
      \item rows $i,i+1,\dots,i+m_1-1$ and columns $j+k+1,j+k+2,\dots,j+m_2-1$
      \end{itemize}
\end{enumerate}

\noindent Stopping rules for the forward selection process will be discussed later after we have dealt with the pruning process.

\subsection{Pruning final scanned regions that are insignificant}

The aim of pruning is to trim away all insignificant offspring and parents. If no scanned region survives the pruning process, then there is no signalled outbreak. Otherwise scanned regions that survive the pruning process highlight the outbreak region (as in Figure 3 and 4 covered later in the application). The FSS plan steps are now defined in more detail.\\
The process of pruning is very simple. We prune the node if
$$\sqrt{\tilde{Y}_{ijt}}-\sqrt{\tilde{M}_{ijt}}<h(\tilde{M}_{ijt})$$
where these $h(.)$ values are positive constants chosen to deliver a specified in-control average run length (ARL).  An estimate of $h$ is determined using a parametric bootstrap for in-control data similar to that defined in this paper but this time the most unusual partition process is used.  Thus the bootstrap process is repeated - now to estimate $h-$ value(s).

\subsection{Stopping rules for recursive partitioning}

 In this section we specify the stopping rule for the partitioning process.  This is based on avoiding computational effort by stopping splitting parent regions when there is no chance of future generations surviving the pruning process. That is, stop whenever
$$\sqrt{\tilde{Y}_{ijt}}<h(\min(\tilde{M}_{ijt})).$$
where $\min(\tilde{M}_{ijt})$ is the minimum cell mean for all cells in the parent space. This stops the tree growing when it is known that the offspring will not survive the pruning process.  The pruning process only leaves generations with smoothed counts significantly higher than expected (see example and Figure 1).
 The last stopping rule is to cease partitioning the parent space when there is no offspring more unusual than the parent(i.e., $\sqrt{\tilde{Y}_{ijt}}-\sqrt{\tilde{M}_{ijt}}$ cannot be increased by partitioning).  In Figure 1(d) this would terminate further generation of offspring for two extra regions (the outbreak region and its left-hand neighbour).

\subsection{Comparing strength and weaknesses}

\paragraph{Bias:}
Biases in variable selection methods are discussed in Miller (1984).  The bias that is common to variable selection in regression modelling and selecting the scanned area is the {\bf competition bias.} In the scanning context this is translated as the bias in selecting between different $m_1 \times m_2$ regions in the scan plan for the outbreak in forward selection process. The FSS plan is likely to reduce the competition bias on the "all-subset" scan plan (see Miller (1984) - forward selection reduces competition biases when compared to all subset explanatory variable selection in regression).\\
The scan statistic in this paper suffers a {\bf size bias} by selecting the size of $m_1$ and $m_2$ scanned region and a temporal memory bias by selecting time window $T$.   The FSS plan suffers very little from a size bias although it does involve a smoothing bias by selecting $\lambda$, which is expected to have a minor influence. It would suffer a temporal memory biases through $\alpha$ but this is expected to be less than the moving window bias of the scan plan.\\
Finally, the scan statistic has a {\bf boundary bias} that will be illustrated in the simulation study section of the paper.\\
Both plans suffer a {\bf shape bias} favouring rectangular regions.

\paragraph{Computation effort:} The number of regions that the Scan plan investigates (assuming a constant scanning region) can be directly determined using the parameters defined for the Scan to proceed. Indeed, with a $A \times B$ region and $m_1 \times m_2$ as the Scan parameters, the total number of region will be equal to $(A - m_1 + 1) \times (B - m_2 + 1)$.
The FSS plan, on the other hand, with $T$ generations considers the following number of regions:
\begin{enumerate}
  \item The first generation considers $ A + B - 2$ partitions ($A-1$ rows partitions and $B-1$ columns partitions of two offspring).
  \item After this the number of partitions to consider diminishes with each generation.  The number of partitions in the worst case without invoking the stopping rule is $ T \times (A + B - 2)$ (and $2 \times T \times (A + B - 2)$ if both offspring are considered) which is less than the number the scan plan above needed to examine.  However in most cases when stopping rules are applied the number is much less than this.
  \end{enumerate}
So the FSS plan needs less computational effort and offers greater flexibility in terms of detecting the unknown variable size outbreaks than the scan plan with a fixed scanning region.

\subsection{Simple Example}

The worked example below is used to demonstrate a simple version of the partitioning process for the FSS plan (when counts are not spatio-temporal EWMA smoothed).  This example used in-control Poisson counts data generated with cell means of 3 for all $10 \times 10$ lattice (see Figure 1), but the red values in Figure 1 are outbreak counts.  Additional outbreak counts were generated for the outbreak cells and they were added to the in-control counts.   The outbreak cell counts have a mean of 6. We assume no knowledge of this outbreak region, and therefore below all cell counts are assumed to have mean 3.  To establish the best row and column partition of a parent region, it is convenient to work with row and column totals of the parent region.   The forward selection approach of the FSS plan is now demonstrated below:

 \noindent{\bf Generation 1:}  Calculate row and column totals for the full matrix in Figure 1(a).  These are
 $$\left(
   \begin{array}{cccccccccc}
      33 &49& 37 &55 &23 &24 &24 &22 &25 &30\\
   \end{array}
 \right)$$ and
 $$\left(
   \begin{array}{cccccccccc}
 26& 35 &32 &38 &34 &34 &33 &41 &28 &21\\
 \end{array}
 \right)$$
   respectively. The row partition that most departs from expected corresponds to column counts $33+49+37+55=174$ with expected value of $120$.  The p-value for this count is $${\rm P}(Y_{11t}\geq 174 \vert m_1=4,m_2=10,M_{11t}=120)=0.0000015.$$  The column partition that departs most from expected is
  $26+35+32+38+34+34+33+41=273$ with expected value of $240$ giving a p-value of $${\rm P}(Y_{51t} \geq 273 \vert m_1=6,m_2=10,M_{11t}=240)=0.0167785.$$
  Clearly the row partition is more unusual, and therefore the first partition is demonstrated in Figure 1 (a).

  \noindent{\bf Generation 2:}  Let offspring 1 be the region above the partition in Figure 1(a).    The row and column totals for this parent space are:

  $$\left(
    \begin{array}{cccc}
    33& 49& 37& 55\\
    \end{array}
  \right)$$
  and
  $$\left(
    \begin{array}{cccccccccc}
    16&19&13 &22& 18 &21 &21 &24& 11 & 9 \\
    \end{array}
  \right)$$

   \noindent respectively. The row partition that most departs from expected corresponds to column counts $49+37+55=141$ with expected value of $90$ giving a p-value of $0.0000003$.  The column partition that departs most from expected is
  $16+19+13+22+18+21+21+24=154$ with expected value of $96$ giving a p-value of $0$.
  Clearly the column partition is more unusual, and therefore the next partition is demonstrated in Figure 1 (b).

  Now looking at offspring 2 from the first generation; the row and column totals for this are:
 $$ \left(
    \begin{array}{cccccc}
  23 &24 &24 &22 &25 &30\\
    \end{array}
  \right)$$
   and
   $$\left(
     \begin{array}{cccccccccc}
   10 &16 &19 &16 &16 &13 &12 &17 &17 &12 \\
     \end{array}
   \right)$$

   \noindent respectively.  The row partition that departs most from expected corresponds to column total 30 with expected value of 30 producing a p-value of $0.452$.  The column partition that departs most from expected on the high side is $17+17+12=46$ with expected value of $54$ producing a p-value of $0.847$.  The row partition is better (see Figure 1(b)).

   \noindent{\bf Generation 3:} Let offspring 1, 2, 3 and 4 from generation 2 be the top left partitioned region,  top right region, the region second from the bottom and bottom regions in Figure 1(b), respectively.

  Now looking at offspring 1, the row and column totals for this are:
  $$\left(
    \begin{array}{cccc}
    27 & 45 & 32 & 50\\
    \end{array}
  \right)$$
   and
   $$\left(
     \begin{array}{cccccccccc}
     16&19&13 &22& 18 &21 &21 &24 \\
     \end{array}
   \right)$$
respectively.  The row partition that departs most from expected corresponds to column total $45+32+50=127$ with expected value of $72$ producing a p-value of $0$.  The row partition that departs most from expected on the high side is $22+ 18 +21 +21 +24=106$ with expected value of $60$ and corresponding p-value equal to $0$.  Both row and column partitions have p-value equal to $0$, but comparing their standardised scores $(127-72)/\sqrt{72}= 6.48$ and $(106-60)/\sqrt{60}=5.94$, the row partition is preferred (see Figure 1(c)).

 Now looking at offspring 2, the row and column totals for this are:
  $$\left(
    \begin{array}{cccc}
    6 & 4 & 5 & 5\\
    \end{array}
  \right)$$
   and
   $$\left(
     \begin{array}{cc}
     11&9 \\
     \end{array}
   \right)$$
respectively.  The row partition that departs most from expected corresponds to column total $6$ with and expected value of $6$ producing a p-value of $0.39$.  The column partition that departs most from expected on the high side is $11$ with an expected value of $12$ and corresponding p-value $0.538$.  The row partition is preferred (see Figure 1(c)).

   Now looking at offspring 3, the row and column totals for this are:
   $$\left(
     \begin{array}{ccccc}
      23 &24 &24 &22 &25 \\
     \end{array}
   \right)$$
   and
   $$\left(
     \begin{array}{cccccccccc}
     8&13&13&16&10&13&10&15&12& 8 \\
     \end{array}
   \right)$$
  respectively.  The column partition that departs most from expected corresponds to column total $25$ with expected value of $30$ producing a p-value of $0.792$.  The row partition that departs most from expected on the high side is $8 +13+ 13 +16=50$ with expected value of $60$ and corresponding p-value of $0.892$.  The column partition is better (see Figure 1(c)).

   Offspring 4 can only partition on columns and the partition that most departs from expected corresponds to counts $4+5=9$ with expected value $6$ and p-value $0.083924$.  (see Figure 1(c))

   \noindent{\bf Generation 4:} The offspring from generation 3 region that contains the outbreak (see Figure 1(c)) has row and column totals equal to:
  $$ \left(
     \begin{array}{ccc}
     49& 37 &55\\
     \end{array}
   \right)$$
   and
   $$\left(
     \begin{array}{cccccccc}
     12&12&11&19&17&17&19&20\\
     \end{array}
   \right)$$
   respectively.  The column partition that departs most from expected corresponds to column total $37+55=92$ with expected value of $72$ producing a p-value of $0.0000475$.  The row partition that departs most from expected on the high side is $19+17+17+19+20=92$ with expected value of $45$ and corresponding p-value of $0$.  However, several partitions have p-values of $0$.  The most unusual row partitions is therefore determined by the split with the highest signal-to-noise ratio.  This is corresponds to $(92-45)/\sqrt{45}=7.01$.  The row partition is better (see Figure 1(d)).

   The remaining offspring for generation 4 are reported in Figure 1(d).  Note that the offspring in the top row and columns 9 and 10 in Figure 1(c) did not generate offspring in the 4th generation. This was because there was no partition more unusual than the parent.

\begin{figure}
  \centering
  \includegraphics[trim=2.5cm 7.5cm 2.5cm 5.5cm, clip=true,scale=0.85]{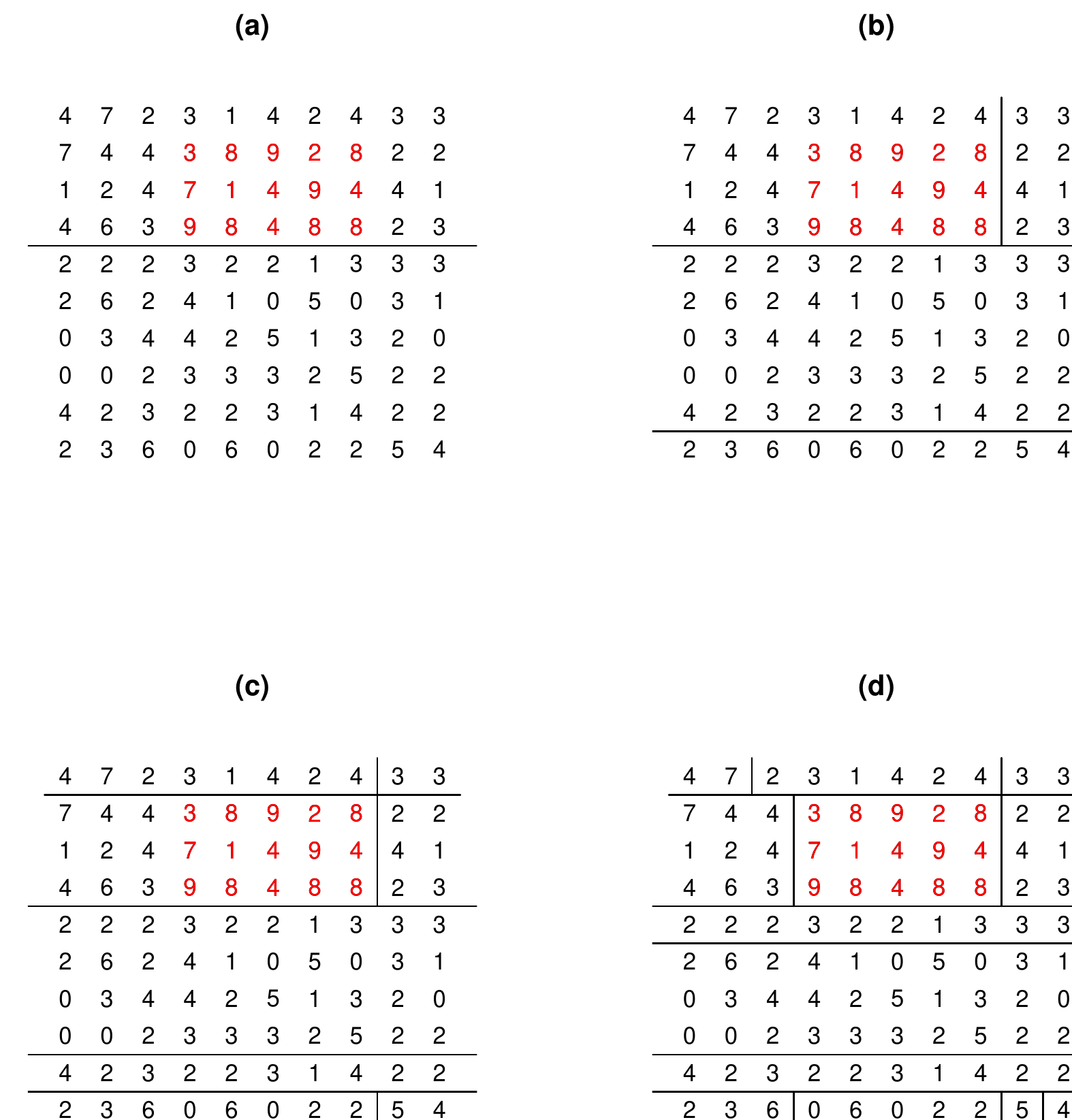}\\
  \caption{Recursive partitioning for an example of in-control cell counts with mean 3 and out-of-control counts in red with mean 6}
\end{figure}

\paragraph{Stopping Rule:} for the example in Figure 1, we assume that $h(M)=3.5-0.0065M$, then we would stop partition when the parent counts are less than or equal to 12.   Partitioning is terminated in  the two smallest offspring regions in (c). A further 5 offspring regions would terminate partitioning in (d).  The used stopping rule substantially reduces the computation effort required for scanning of all potential regions using the traditional scan statistic.  If we started pruning from generation 4 in Figure 1 using $h(M)=3.5-0.0065M$, then only one offspring containing the outbreak regions would survive the pruning process.

We also stop at the $6^{th}$ generation because this is sufficient for detecting multiple outbreak regions.   This rule will generally detect three unconnected outbreak region that are rectangular.  If there are less than 6 outbreak regions then the plan with 6 generations should diagnose all of them, particularly if some are close enough to be grouped into a single outbreak.  An advantage of stopping after 6 generations is the reduction in competition biases discussed earlier, but it mostly helps in reducing the computational effort.

\section{Simulation study}

In this application, the first two years of data are used to estimate the bootstrap population for the Poisson cell counts.  This is done by fitting a Poisson regression model to the first 2 years of cell counts, separately for each cell. These models are then used to generate a bootstrap time series of cell counts (called bootstrap samples).
The fitted models using the bootstrap approach for the simulation data discussed later in Section 4 is:
          \vskip 3pt
             $ \begin{array}{l} \\
              ${\rm E}(X)$=0.0192+0.784214z_p+0.007253 n_s-0.0001576 n_s^2-0.000219 \mu+\\
                                       0.00020 \mu_p+0.420963 z_p^2-0.010587 z_p \, n_s+0.000159z_p \, n_s^2-\\
                                       0.000341 z_p \, \mu+0.0003577z_p \, \mu_p-0.000378 \mu \, z_p^2+0.000144 \mu_p \, z_p^2 \\
              \end{array}$
              \vskip 3pt
              and
              \vskip 3pt
              $\begin{array}{l}\\
${\rm Var}(X)$=exp(-3.3696-1.844822 z_p+0.03334 n_s-0.000904 \mu+\\
           0.000792 \mu_p+0.736898z_p^2+0.057413 z_p\, n_s-0.001137 z_p \,\mu+\\
           0.001294 z_p \, \mu_p-0.007494 z_p^2\, n_s+0.000378 \mu \, z_p^2+0.000489
            \mu_p \, z_p^2 ).\\
\end{array}$
\vskip 3pt
\noindent The most unusual partition is the one that maximizes $(X-{\rm E}(X))/\sqrt{{\rm Var}(X)}$.  We threshold $\sqrt{{\rm Var}(X)}$ to not fall below $0.25$ to avoid partitioning on cells with low departures from expected.

The scan statistic with $m_1=m_2=10$ is used.  This scans $1/16$ of the rectangular cells in the target area ($A=40$ by $B=40$ cells), i.e., the plan is designed to target fairly large “square” clustered outbreaks.  The outbreak size is generated as rectangles involving $40$ cells.   The scan plan searches all possible $10$ by $10$ regions.   The simulated counts were generated homogeneous mean counts with each cells assumed to have a mean of $0.01$ for all days, and the FSS plan used $\alpha=0.1$ and $\lambda=0.7$ .  All plans are designed to have an in-control ARL equal to $100$.
The simulation process generated in-control counts to determine the FSS partitioning process and pruning rule, and to determine $h_{scan}$ for the scan statistic. The plans' detection performance for out-of-control situations were simulated by adding additional generated counts for a fixed rectangular outbreak region.   These extra counts were added to the in-control counts.  The outbreak region is then hidden and we examined how early the plans alarm the hidden outbreaks.  Rectangular outbreak regions were generated involving $10$ by $4$ cells and $20$ by $2$ cells.   The ARLs results reported in Table 1 are estimated from $1000$ simulations. The FSS plan’s properties are unknown and its flexibility in partitions appears to indicate its strength in a robust performance across a range of outbreak dimensions.  In other words, if the outbreak dimensions are known, then the Scan plan can always be trained to be more efficient than the FSS plan, and therefore it is preferred in these circumstances.  However, the FSS plan could also be trained to exploit this information and therefore it is an open research question as to which plan offers computational more efficient results even in this unrealistic situation.

To keep things simple, we used $h(M)=1.3$ for the FSS plan in the simulation because stopping the plan at the $6$th generation did not produce an excessive number of offspring with small mean counts as in Sparks and Okugami(2010).  Therefore the need for $h(.)$ to be a function of the child mean was avoided by this early stopping rule.   Slight improvements to large outbreaks ($50\%$ or more of the target region) could be established using a model of the form of $h(M)=a+bM$.  Estimating this threshold is a little more complicated and a process for doing this is outlined in Sparks and Okugami (2010).  The Scan plan using $h_{scan}$ are defined by the p-value was inadequate because its value often converged too quickly to 0.  Therefore the traditional standardised statistic was used to find outbreaks and this threshold was $7.809$.

In this example, the simulation process used an in-control mean of $0.01$ for all cells and out-of-control means were taken as $0.01 \times (1+\delta)$. The corresponding $h(M)$ value used is $1.3$.
In Table 1, the plan with the lowest ARL for a generated outbreak is reported in bold text (the better plan) making it easy to see trends in performance.  Note that the FSS plan is better at finding the outbreaks on the boundary than the scan plan.  Also note that the FSS plan is less influenced by the position of the outbreak - as long as the area and step change are the same the FSS plan's the out-of-control ARLs are similar. For $10$ by $4$ outbreaks, the scan plans signals smaller outbreaks ($\delta<3$) on average earlier than the FSS plan.  There are substantial advantages in using the FSS plan when the outbreak is close to the boundary or very different in shape from the scanned area $m_1$ by $m_2$ for all but small step changes. The scan statistic is less likely to signal an outbreak of the same size situated on the boundary (thus confirming the boundary bias eluded to earlier in the paper).

The in-control recurrence intervals (see Fraker et al 2008) for these plans were very similar,i.e., $84.3$ and $87.3$ for the FSS and Scan plan, respectively.

\vskip 6pt

Table 1: ARL performance of the plans when the in-control $ARL=100$ and the outbreak spans a region of $10$ by $4$ or $20$ by $2$ pixels.  The Scan plan uses $T=10$ and $m=10$.

\vskip 3pt
  \begin{tabular}{|c|rr|rr|rr|rr|}\hline
  Method&FSS&Scan&FSS&Scan&FSS&Scan&FSS&Scan\\
     \hline
     &\multicolumn{8}{|c|}{Outbreak region}\\  \hline
     Row&\multicolumn{2}{|c|}{5:14}&\multicolumn{2}{|c|}{10:19}&\multicolumn{2}{|c|}{1:20}&\multicolumn{2}{|c|}{1:20}\\
     Column&\multicolumn{2}{|c|}{11:14}&\multicolumn{2}{|c|}{16:19}&\multicolumn{2}{|c|}{10:11}&\multicolumn{2}{|c|}{1:2}\\
     $\delta$&\multicolumn{8}{|c|}\hrulefill \\
     0.0& 100.2 & 100.5 & 100.2 & 100.5 & 100.2 & 100.5&100.2 & 100.5\\
     0.5 & 27.03 & {\bf 13.65} & 26.45 & {\bf 13.91} &{\bf 25.89}&27.18&{\bf 24.59}&39.96\\
     1.0 & 8.89 & {\bf 6.52} & 8.66 & {\bf 6.70} & {\bf 8.68}& 8.79& {\bf 8.79}&12.79\\
     2.0 & {\bf 3.49} & 3.55 & {\bf 3.69} & {\bf 3.69} &{\bf 3.59}&5.33&{\bf 3.45}&6.42\\
     3.0 & {\bf 2.38} & 2.56 & {\bf 2.43} & 2.62&{\bf 2.28}&3.65&{\bf 2.29}&4.53 \\
     4.0 & {\bf 1.89} & 2.14 & {\bf 1.99} & 2.09&{\bf 1.82}&2.84&{\bf 1.82}&3.43 \\
     5.0 & {\bf 1.55} & 1.84 & {\bf 1.59} & 1.76 &{\bf 1.54}&2.42&{\bf 1.54}&3.05\\
     6.0 & {\bf 1.35} & 1.62 & {\bf 1.42} & 1.66 &{\bf 1.36}&2.02&{\bf 1.33}&2.40\\
     7.0 & {\bf 1.17} & 1.39 & {\bf 1.19} & 1.44 &{\bf 1.15}&1.99&{\bf 1.14}&2.17\\
     8.0 & {\bf 1.09} & 1.22 & {\bf 1.10} & 1.17 &{\bf 1.15}&1.87&{\bf 1.10}&1.96\\
     9.0 & {\bf 1.00} & 1.09 & {\bf 1.01} & 1.07&{\bf 1.03} &1.51&{\bf 1.00}&1.65\\
     \hline
   \end{tabular}

\vskip 6pt

\paragraph{Computation effort:} in this simulation study we consider $A=B=40$ and the scan plan with $m_1=m_2=10$.  In this case the scan plan investigates $31 \times 31=961$ regions in search for an outbreak.  The FSS plan (6 generations) considers the following number of regions:
\begin{enumerate}
  \item The first generation considers 78 partitions (39 rows partitions and 39 column partitions of two offspring).
  \item After this the number of partitions to consider diminishes with each generation.  The number of partitions in the worst case without invoking the stopping rule is $5 \times 78 $ 
  \end{enumerate}
The total number of regions considered is then $468$ which is smaller than $961$ (and this is without considering the other dimensions, such as time, etc).

\section{Example of application}

The AEGISS data (Diggle {\emph{et al.} 2003) contains space-time locations
for 10,572 cases of non-specific gastrointestinal disease in the county of Hampshire, UK. Each ($x$,$y$)-location corresponds to the centroid of the unit post-code of the residential address of the person with the disease. The unit of distance is 1 metre. The unit of
time is in days (the date the disease is reported).  The region is divided into a $40$ by $40$ lattice structure such that their marginal row and column disease counts are equal (see Figure 2 where all incidents are plotted and the lattice structure is plotted).  The data spanned $1095$ days which means that there are just under $10$ incidents per day.  With there being $1600$ cells in the lattice this gives roughly an expected value of $0.006$ per cell.  Since the Hampshire county is not rectangular there are a number of boundary cells with zero incidents (See Figure 2 with all incidents and the lattice design) therefore the cell means for cells with counts are roughly 0.01 on average.  The empty cells had their expected counts set equal to $0$.  For each cell in the lattice, their first $730$ days were used to fit a Poisson regression model to the cell counts using day-of-the-week as a factor, harmonics in time as explanatory variables.  This fitted model was then used to forecast the expected cell counts for day $731$.  Cells with very low counts where grouped with neighbours for modelling and therefore were forecasted at the aggregated group level.  These forecasts were then proportional distributed to the cells that were combined for modelling.  Forecasts were started using training data (first $2$ years of data) and then updated daily using a moving window of 730 days.  The actual counts were compared to their forecasts (i.e., expected values) and Figure 3 presents the results of the FSS plan.  The dashed lines indicate the scanned region that proved significant after pruning.

\begin{figure}
  \centering
  \includegraphics[trim=2.2cm 5.5cm 2.2cm 6.5cm, clip=true,width = 9cm]{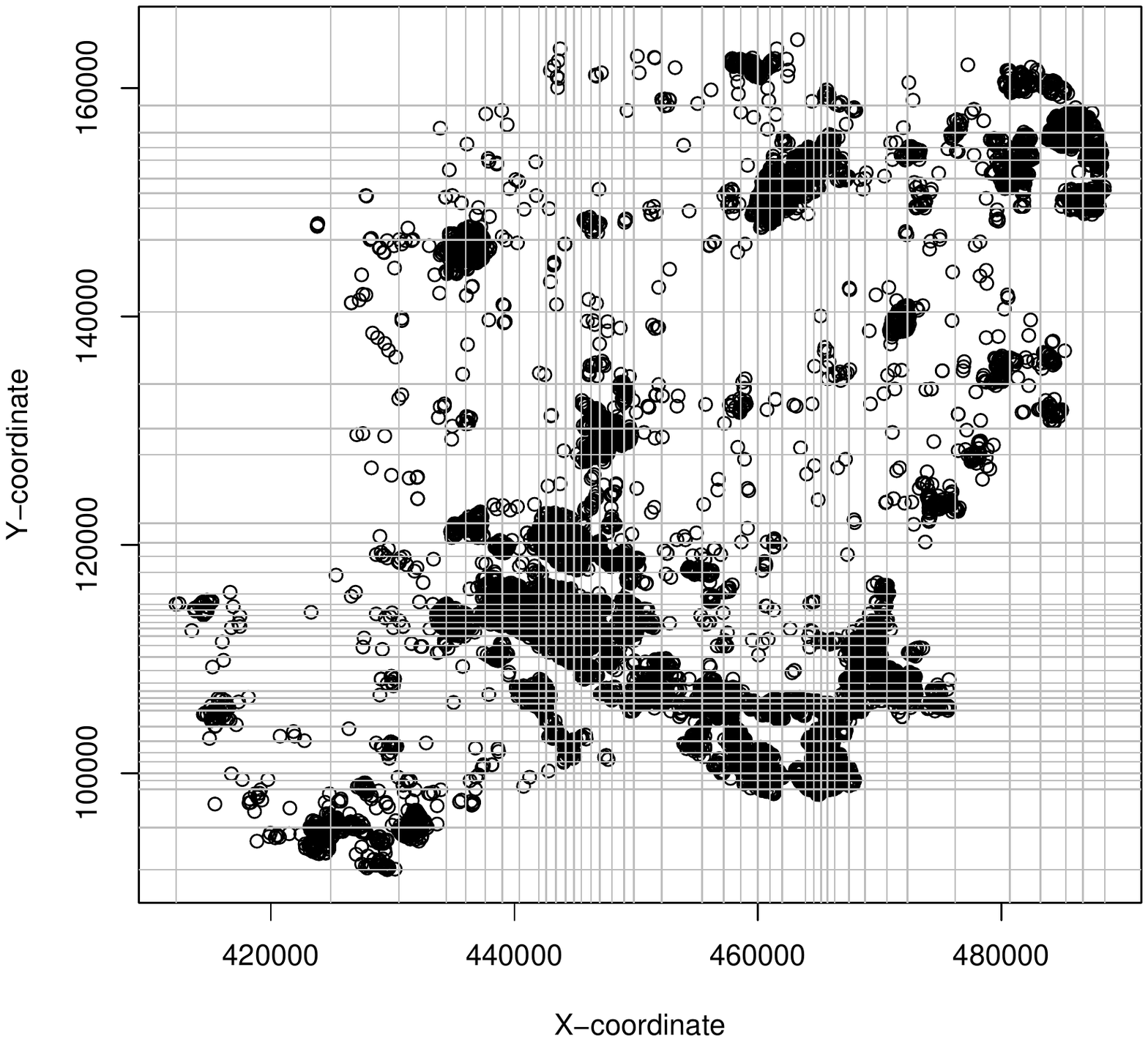}\\
  \caption{AEGISS data and lattice design}
\end{figure}

On day $731$ (in Figure 3, left hand side) note that there were $21$ incidents of gastrointestinal disease reported, while on days $730$ and $729$ there were $12$ and $10$ (just above the average number of incidents per day), respectively.  The total number of incidents for Day $731$ is unusual.  The FSS plan suggests (dashed rectangles in Figure 3, lhs) that this outbreak relative to forecasts is confined to the far North and South-East of Hampshire.

\begin{figure}
  \centering
  \includegraphics[trim=2.3cm 5.5cm 2.2cm 5.8cm, clip=true,width=7.5cm]{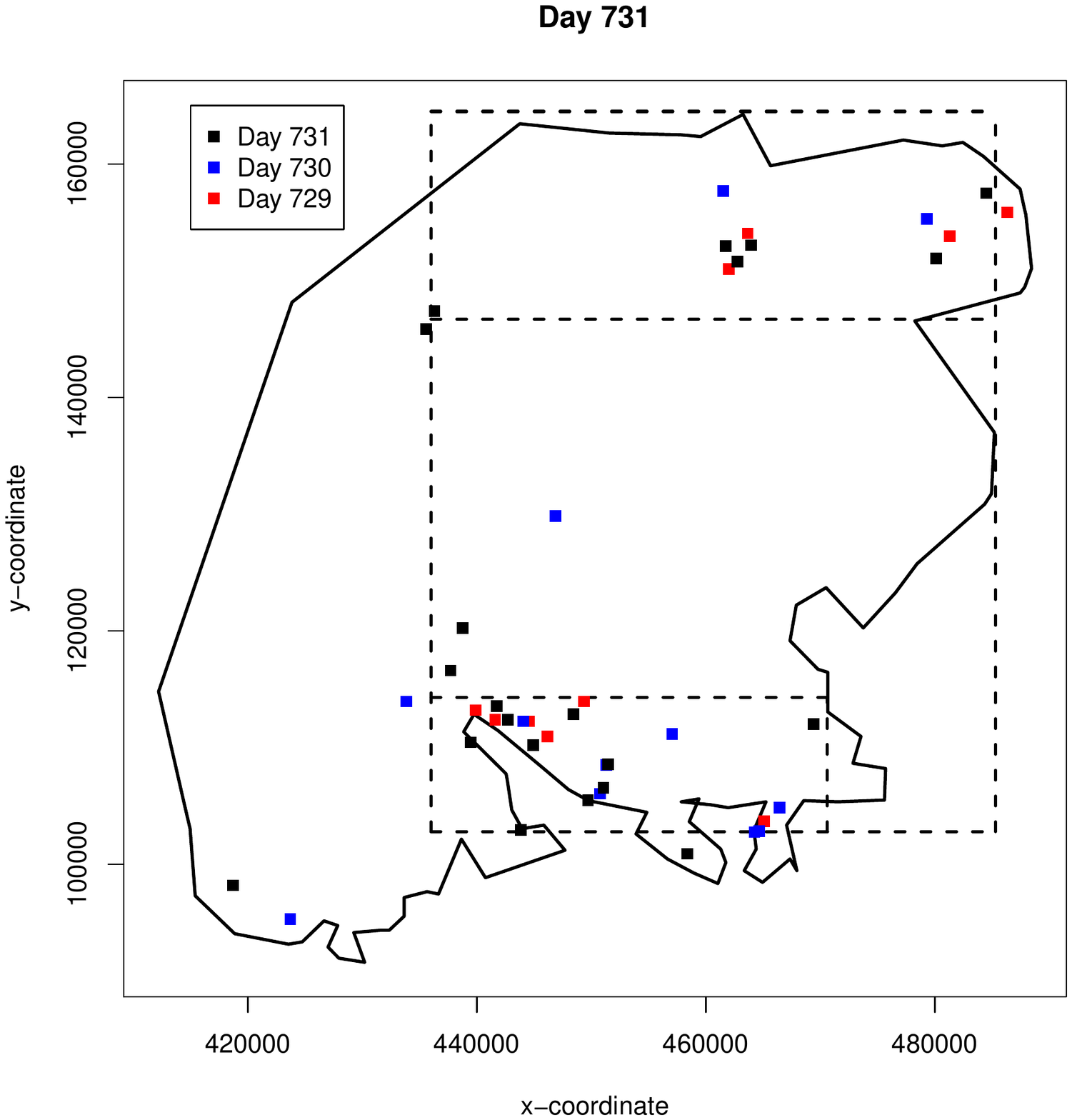}
\includegraphics[trim=2.3cm 5.5cm 2.2cm 5.8cm, clip=true,width=7.5cm]{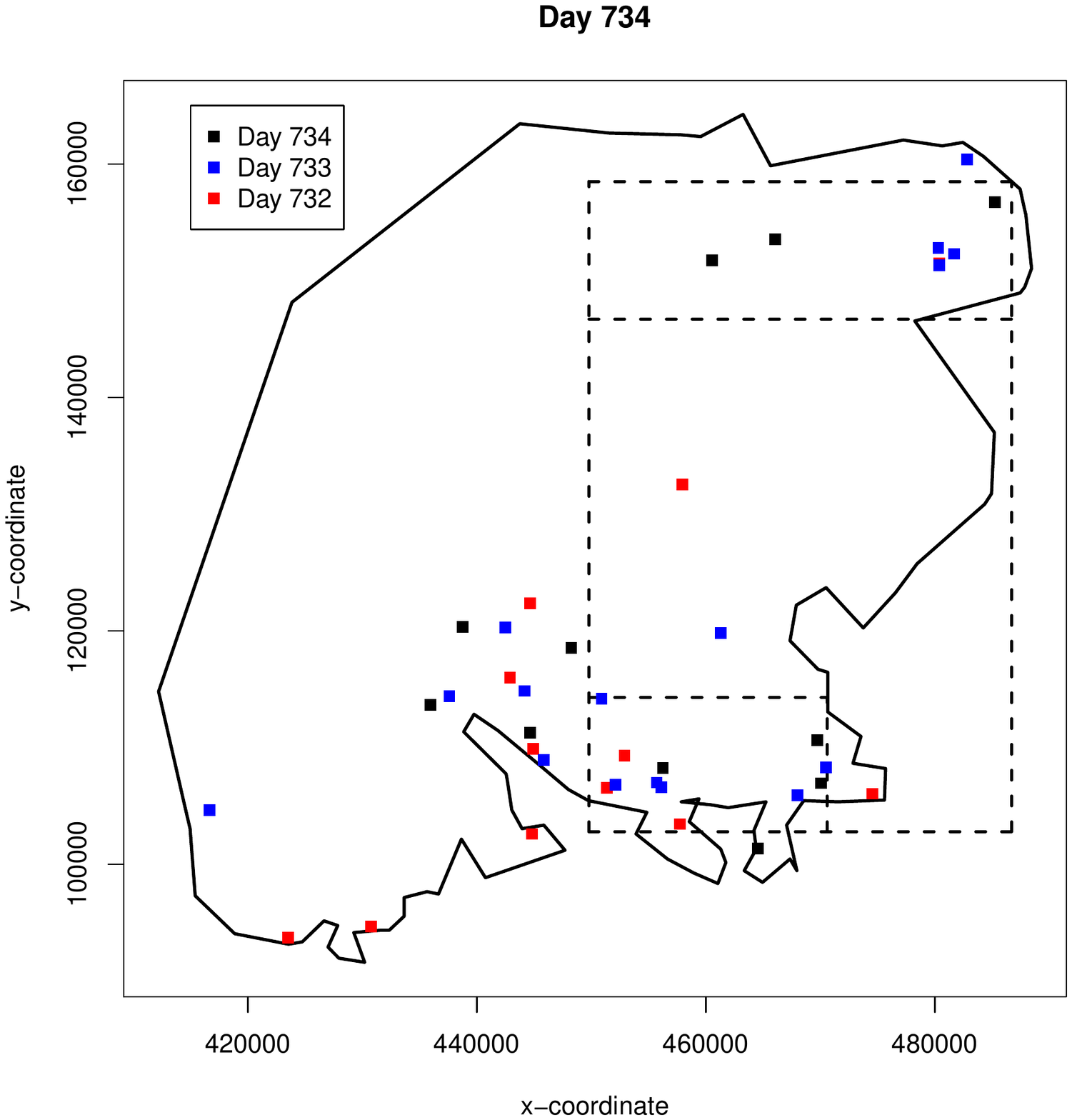}
  \caption{Outbreak detected in the AEGISS data}
\end{figure}

Figure 3 (right hand side) looks at the last day that this outbreak signalled (Day $734$).  Note that Day $734$ had $11$ incidents, Day $733$ had $16$ incidents and Day $732$ had $12$ incidents (all above the average per day).  Day $735$ had $6$ incidents (well below the average) so this day and the next week does not signal indicating the end of this outbreak.  The number of incidents start reducing on Day $734$, and note from Figure 3 (rhs) that the signalled outbreak region also starts shrinking further East both in the South and the North of Hampshire.

\noindent If the update forecasts using a moving window of $730$ days, other persistent outbreaks are signalled and these can be obtained from the author on request.

\section{Concluding remarks}

If the shape and size of future outbreaks are known, then the scan plan can be designed to detect these
early.  However, in cases where nothing is known about future outbreaks, the FSS plan offers an effective, robust and computationally efficient outbreak detection methodology.  Its relative performance seems to improve as the lattice structure reduces from 100 by 100 cells to 40 by 40 cells (results of simulations not reported in this paper).  The best advantage of recursive partitioning is that it can be easily scaled up to higher dimensions of age, gender, work address, etc. (see Sparks and Okugami, 2010), whereas the Scan plan becomes unworkable for any more than three dimensions.  For example, if we found that an outbreak related peoples' work geographical location at A and living at location B (different from A), then this may indicate the transmission location as the public transport services between these two locations.

The FSS plan is similar to forward selection in model selection which is feasible when the all subset selection methods becomes unworkable for very large set of explanatory variables (Miller, 1984). Similarly the FSS plans has it biggest advantage when search for outbreaks in more than three dimensions or over a large number of cells in the target region.

A further work of interest would be to consider replacing use of the lattice in the FSS plan with a modified Ripley's correction for spatial processes (see Charpentier and Gallic, 2013). This correction aims to remove biases on the boundary of the target region.  Testing how this could be used to improve the FSS plan is left as a future research topic.

\section*{References}

\begin{enumerate}
\item Charpentier, A., Gallic, E. (2013). Visualizing spatial processes using Ripley's correction: an application to bodily-injury car accident location, Technical Report CREM, {\it http://hal.archives-ouvertes.fr/hal-00725090}.
\item Chen, J., and Glaz, J. (1996). Two-Dimensional Discrete Scan Statistics, {\it Statistics and Probability Letters}, 31, 59-68.
\item Diggle, P, Knorr-Held, L, Rowlingson, B, Su, T, Hawtin, P and Bryant, T (2003). On-line monitoring of public health surveillance  data. In {\it Monitoring the Health of Populations: Statistical Principles and Methods for Public Health Surveillance} editors R. Brookmeyer and D.F. Stroup, pages 233-66. Oxford : Oxford University Press.
\item Fraker SE, Woodall WH, Mousavi S.(2008). Performance metrics for surveillance schemes. {\it Quality Engineering}, 20:451-464.
\item Glaz, J., J. Naus, and S. Wallenstein (2001) {\it SCAN Statistics}. Springer, New York.
\item Grigg, O., Spiegelhalter, D. (2007). A simple risk-adjusted exponentially weighted moving average, {\it Journal of the American Statistical Association}, 102, 140-152.
\item Han, S. W., Y. Mei, and K. -L. Tsui (2008). A comparison between SCAN and CUSUM methods for detecting increases in Poisson rates, {\it Technical Report}, School of ISyE, Georgia Institute of Technology.
\item Kulldorff M and Nagarwalla N. (1995). Spatial disease clusters: Detection and Inference. {\it Statistics in Medicine}, 14:799-810.
\item Kulldorff M. (1997). A spatial SCAN statistic, {\it Communications in Statistics: Theory and Methods}, 26:1481-1496.
\item Kulldorff, M. (2001). Prospective time periodic geographical disease surveillance using a SCAN statistic, {\it J.R. Statist. Soc. A.}, 164: 61-72.
\item Kulldorff M, Heffernan R, Hartman J, Assunção RM, Mostashari F. (2005).  A space-time permutation SCAN statistic for the early detection of disease outbreaks, {\it PLoS Medicine}, 2:216-224.
 \item Lowry, C.A., Woodall, W.H., Champ, C.W., Rigdon, S.E. (1992): A multivariate exponentially weighted moving average control chart. {\it Technometrics}, 34, 46-53.
\item Miller, A.J. (1984). Selection of Subsets of Regression Variables. {\it J.R. Statist. Soc. A.} 147(3):389-425.
\item Raubertas, RF (1989). An analysis of disease surveillance data that uses the geographic locations of reporting units, {\it Statistics in Medicine}, 18, 2111-2122.
\item Rogerson, P.A. and I. Yamada. Monitoring Change in Spatial Patterns of Disease: Comparing Univariate and Multivariate Cumulative Sum Approaches. {\it Statistics in Medicine}, 23 (14), 2004, 2195-2214.
\item Sparks, R. (1992). Quality control with multivariate data.  {\it Australian Journal of Statistics}. 34(3):375-390.
\item Sparks, R, Carter,C, Graham, PL, Muscatello, D, Churches, T, Kaldor, J, Turner, R, Zheng, W and Ryan, L.  (2010). Understanding sources of variation in syndromic surveillance for early warning of natural or intentional disease outbreaks.  {\it IIE Transactions}, 42(9): 613-631.
\item Sparks, R.S and Okugami, C (2010). Surveillance trees: early detection of unusually high number of vehicle crashes, {\it InterStat}, January,  see http://interstat.statjournals.net/YEAR/2010/abstracts/1001002.php
\item Takahashi, K., Kulldorff, M., Tango, T., and Yih, K. (2008). A flexibly shaped space-time scan statistic for disease outbreak detection and monitoring, {\it Journal of Health Geographics}. 7:14.
\item Woodall, W. H., Marshall, J. B., Joner, M. D., Jr., Fraker, J. E., and Abdel-Salam, A. G.  (2008). On the use and evaluation of prospective SCAN methods for health-related surveillance, {\it J.R. Statist. Soc. A.}, 171: 223-237.
\end{enumerate}

\end{document}